\documentclass[aps,prl,twocolumn,superscriptaddress]{revtex4-1}
\usepackage{graphicx}
\usepackage{dcolumn} 
\usepackage{graphics}
\usepackage{amsmath,amsfonts,amssymb}
\usepackage{hyperref}
\usepackage{float}

\begin{document}
\title{Dispersive Gate Sensing the Quantum Capacitance of a Point Contact}
\author{M. C. Jarratt}
\affiliation{ARC Centre of Excellence for Engineered Quantum Systems, School of Physics, The University of Sydney, Sydney, NSW 2006, Australia.} 
\author{A. Jouan}
\affiliation{ARC Centre of Excellence for Engineered Quantum Systems, School of Physics, The University of Sydney, Sydney, NSW 2006, Australia.} 
\author{A. C. Mahoney}
\affiliation{ARC Centre of Excellence for Engineered Quantum Systems, School of Physics, The University of Sydney, Sydney, NSW 2006, Australia.} 
\author{S. J. Waddy}
\affiliation{ARC Centre of Excellence for Engineered Quantum Systems, School of Physics, The University of Sydney, Sydney, NSW 2006, Australia.} 
\author{G.~C.~Gardner}
\affiliation{Department of Physics and Astronomy and Birck Nanotechnology Center, Purdue University, West Lafayette, Indiana 47907, USA}
\affiliation{Microsoft Quantum Purdue and School of Materials Engineering, Purdue University, West Lafayette, Indiana 47907, USA}
\author{S. Fallahi}
\affiliation{Department of Physics and Astronomy and Birck Nanotechnology Center, Purdue University, West Lafayette, Indiana 47907, USA}
\author{M. J. Manfra}
\affiliation{Department of Physics and Astronomy and Birck Nanotechnology Center, Purdue University, West Lafayette, Indiana 47907, USA}
\affiliation{Microsoft Quantum Purdue and School of Materials Engineering, Purdue University, West Lafayette, Indiana 47907, USA}
\affiliation{School of Electrical and Computer Engineering, and School of Materials Engineering, Purdue University, West Lafayette, Indiana 47907, USA}
\author{D. J. Reilly$^{\dagger}$}
\affiliation{ARC Centre of Excellence for Engineered Quantum Systems, School of Physics, The University of Sydney, Sydney, NSW 2006, Australia.} 
\affiliation{Microsoft Quantum Sydney, The University of Sydney, Sydney, NSW 2006, Australia.} 

\begin{abstract}

The technique of dispersive gate sensing (DGS) uses a single electrode to readout a qubit by detecting the change in quantum capacitance due to single electron tunnelling. Here, we extend DGS from the detection of discrete tunnel events to the open regime, where many electrons are transported via partially- or fully-transmitting quantum modes. Comparing DGS with conventional transport shows that the technique can resolve the Van Hove singularities of a one-dimensional ballistic system, and also probe aspects of the potential landscape that are not easily accessed with dc transport. Beyond readout, these results suggest that gate-sensing can also be of use in tuning-up qubits or probing the charge configuration of open quantum devices in the regime where electrons are delocalized. 
\end{abstract}
\maketitle
Fast charge sensors, such as the radio frequency (rf) single electron transistor (rf-SET) \cite{Schoelkopf1238} or rf quantum point contact (rf-QPC) \cite{Reilly:2007}, can rapidly detect the discrete tunnelling of  electrons with a sensitivity close to the limit set by quantum mechanics \cite{Dev_Sch_Nat,Korotkov}.  As such, these devices have proven fundamental in examining the electronic configuration of nanoscale quantum systems and are currently of interest as detectors for reading-out the state of qubits in quantum computers \cite{PhysRevLett.103.160503,Morello}. An alternative to direct charge sensing makes use of the dispersive shift of an $LC$ resonator to probe the quantum capacitance \cite{BUTTIKER1993364,PhysRevLett.68.3088} associated with single-electron tunnelling \cite{Petersson:2010, Schroer:2012, Chorley:2012}. Simplistically, in this regime, a small oscillating voltage $\delta v$ can induce charge $\delta q$ that appears as an additional capacitive contribution to the resonator $\delta c = \delta v / \delta q$, altering its resonance frequency by an amount, $\delta f_0 \sim  -\delta c f_0 / 2C$, where $C$ is the total capacitance. 

From the point of view of scaling-up qubits, the dispersive approach has some advantages over conventional charge sensing in that the capacitance of gate electrodes can be incorporated in the resonator, altogether removing the need for additional  sensors for qubit readout \cite{Colless:2013}. Recent work has demonstrated this technique of dispersive-gate sensing (DGS) to be a highly sensitive probe \cite{Gonzalez-Zalba:2015} of single-electron phenomena \cite{Verduijn:2014, Pei:2015,Scarlino:2017, Mi:2018,Damaz} and capable of enabling high-fidelity single-shot measurements, where the state of a qubit is detected in a single-pass, before it relaxes to the ground-state \cite{West:2018, Rossi:2017}. 

\begin{figure}
	\includegraphics[width=8.6cm]{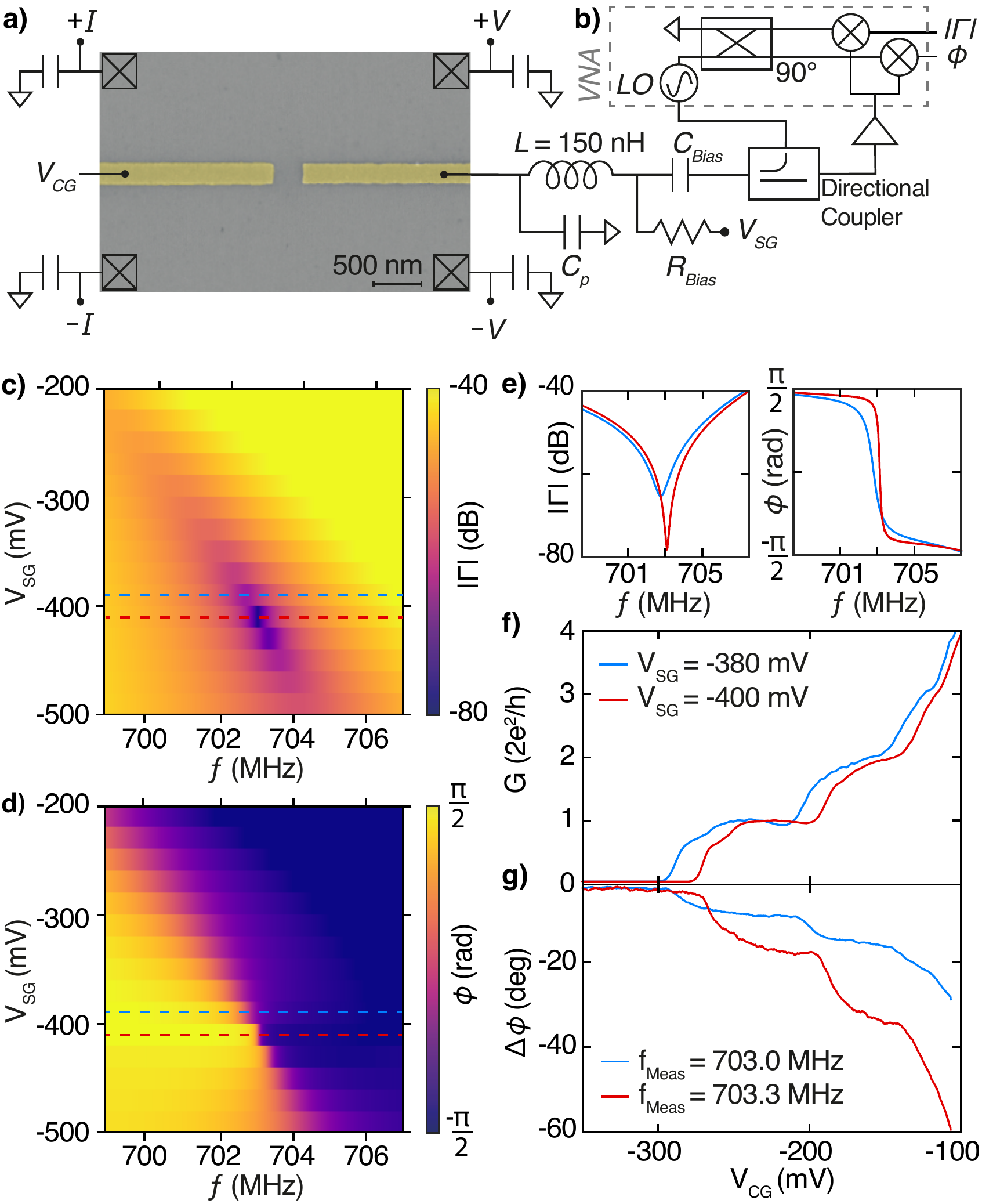}
	\caption{(a) Electron micrograph of the device showing a QPC formed in  between two metallic gate electrodes. (b) The rightmost gate is bonded to an inductor $L$=150 nH, in resonance with the parasitic capacitance of the device, $C_{p}$. A bias tee allows for application of a negative voltage to the gate. After amplification, the demodulated magnitude $|\Gamma|$ and phase $\phi$ responses are read out by a network analyzer. (c) $|\Gamma|$ and (d) $\phi$ as a function of $V_{SG}$ and frequency $f$ near resonance. (e) $|\Gamma|$ and $\phi$ at $V_{SG}$ = -380 mV (blue) and -400 mV (orange), with varying $Q$-factors of $\sim$1100 and $\sim$4200, respectively. (f) Conventional zero-bias conductance measurement for the two values of $V_{SG}$. (g) Change in the demodulated phase response $\Delta \phi$ for the same swept $V_{CG}$ range as (f), taken at $f_{Meas} =$ 703.0 MHz (orange) and 703.3 MHz (blue) to maximise dynamic range. }
\end{figure}

In this paper we extend the dispersive-gate technique beyond the detection of single electron tunnelling, showing that it is also well-suited to probing the quantum capacitance of an open system, where many electrons are transported via partially- or fully-transmitting quantum modes. Similar to compressibility measurements \cite{PhysRevLett.105.256806}, quantum capacitance is a thermodynamic quantity that can reveal fundamental information about the charge configuration of a quantum system \cite{Ilani:2006,Mahan}. For instance, using a QPC formed in a two-dimensional electron gas (2DEG) as a model device, we show that DGS can detect Van Hove singularities associated with the one-dimensional (1D) sub-bands, essentially probing the density-of-states (DOS) in the QPC without the use of a transport current or proximal charge sensor \cite{PhysRevLett.98.196805}. The same DGS approach also reveals the presence of unwanted localized charge-states \cite{Croot:2017} that often form as the QPC is fully-depleted. We compare our DGS measurements with conventional transport, including finite-bias spectroscopy, and relate our experiment to a simple model for the quantum capacitance of an ideal 1D system. 

In comparison to previous measurements of mesoscopic capacitance \cite{Ilani:2006}, the present work leverages the nanoscale geometry of a gate electrode to act as a highly-local probe. We show that the technique remains sensitive to the local DOS in a QPC, even in the open regime where the channel is populated by several sub-bands and screening diminishes the sensitivity of conventional charge detectors.  Such a regime has recently become of interest for reading-out the parity-state of a topological qubit by detecting changes in the hybridization of an electron wavefunction to a quantum dot or reservoir \cite{PhysRevB.95.235305,PhysRevLett.122.016801}. Beyond just detecting state-dependent single electron tunnelling, our work also highlights the potential for DGS to be used in the tune-up of qubit devices, remaining sensitive to charge-states over a range that spans the few-electron tunnelling regime to essentially an open quantum system. The use of DGS to cover this full range of operating conditions then alleviates altogether the need for conventional charge sensors in scaled-up qubit arrays.

Our setup for detecting the quantum capacitance of a QPC is shown in Fig. 1(a). Application of a negative voltage to two TiAu gates on the surface of a GaAs/Al$_{0.3}$Ga$_{0.7}$As heterostructure (electron density 1.33 $\times 10^{11}$ cm$^{-1}$, mobility 3.50 $\times 10^{6}$ cm$^{2}$/V s), forms the QPC in the 2DEG, 91 nm below the surface. The QPC has lithographic dimensions of length 200 nm and width 400 nm. Annealed ohmic contacts, two on either side of the QPC, provide source and drain reservoirs for conventional four-terminal transport measurements, as well as capacitively coupled rf-grounds. The rightmost gate, used for dispersive sensing, is wire bonded to a proximal $LC$ resonator \cite{Hornibrook:2014}, comprizing a superconducting NbTi planar spiral inductor $L$ = 150 nH in resonance with the intrinsic parasitic capacitance of the device, including the unbiased gate, $C_{p} \sim$ 0.342 pF at a frequency of $f_{0} = 1 / 2\pi\sqrt{LC_{p}} \sim 703$ MHz. Our $LC$ resonator is, in-fact, part of a network of on-chip resonators, constructed to enable frequency multiplexing of signals from many gates \cite{Hornibrook:2014}.  An AuPd resistor and parallel-plate capacitor form an on-chip bias-tee, allowing dc voltages to be applied also to the sensing gate $V_{SG}$. The other gate of the QPC is connected only to a voltage source, $V_{CG}$.  The device and resonator-chip are packaged and mounted at the base of a dilution refrigerator with base temperature $T \sim$ 20 mK. The reflected rf response $\Gamma = |\Gamma| e^{i\phi}$ from the sensing gate is amplified at the 4 K stage of the refrigerator and again at room temperature, where it is demodulated to yield amplitude $|\Gamma|$ and phase $\phi$ components using a vector network analyzer. 

Biasing $V_{SG}$ negatively depletes the 2DEG underneath the gate and reduces the 2D geometric capacitive contribution of the resonator [see Fig. 2(d)]. Lowering the capacitance increases the resonance frequency $f_{0}$, [as shown in $|\Gamma|$ in Fig. 1(c) and $\phi$ in (d)], and also changes the quality-factor of the resonator by altering its impedance-match with respect to the characteristic impedance of the transmission line ($Z_{0} =$ 50 $\Omega$). Ideally, we wish to independently adjust $f_{0}$ and the match to the feedline to ensure maximum sensitivity. Similar to approaches that exploit varactors to tune the response of a resonator, \cite{Ares:2015, Ibberson:2018}, we make use of an additional parallel resonator that is part of our frequency multiplexing chip \cite{Hornibrook:2014} to independently tune the matching between the resonator and the feedline. This tunability enables independent control over the $Q$-factor and resonance frequency. Adjusting the phase response in this way, $\phi$ is measured for two different values of $V_{SG}$ in Fig 1. (e), reaching a $Q$-factor of $\sim$ 4200.\\

For each of these two values of $V_{SG}$, a four-terminal measurement of the conductance $G$ through the QPC is performed, as a function of $V_{CG}$. Familiar conductance plateaus, quantized in units of $2e^{2}/h$, are seen in Fig. 1(f). After electrically disconnecting the ohmics from all current and voltage sources, the equivalent sweep is performed for a measurement of $\Delta \phi$, shown in Fig. 1(g). The measurement frequency is chosen to maximise the dynamic range of $\Delta \phi$. Inflections are observed in the phase response of the resonator at gate voltages corresponding to  plateaus in the conductance. We attribute these inflections to the Van Hove singularities in the 1D DOS. Tuning the $Q$ factor of the resonator adjusts the effective gain of the phase response.  \\

\begin{figure}
	\includegraphics[width=8.6cm]{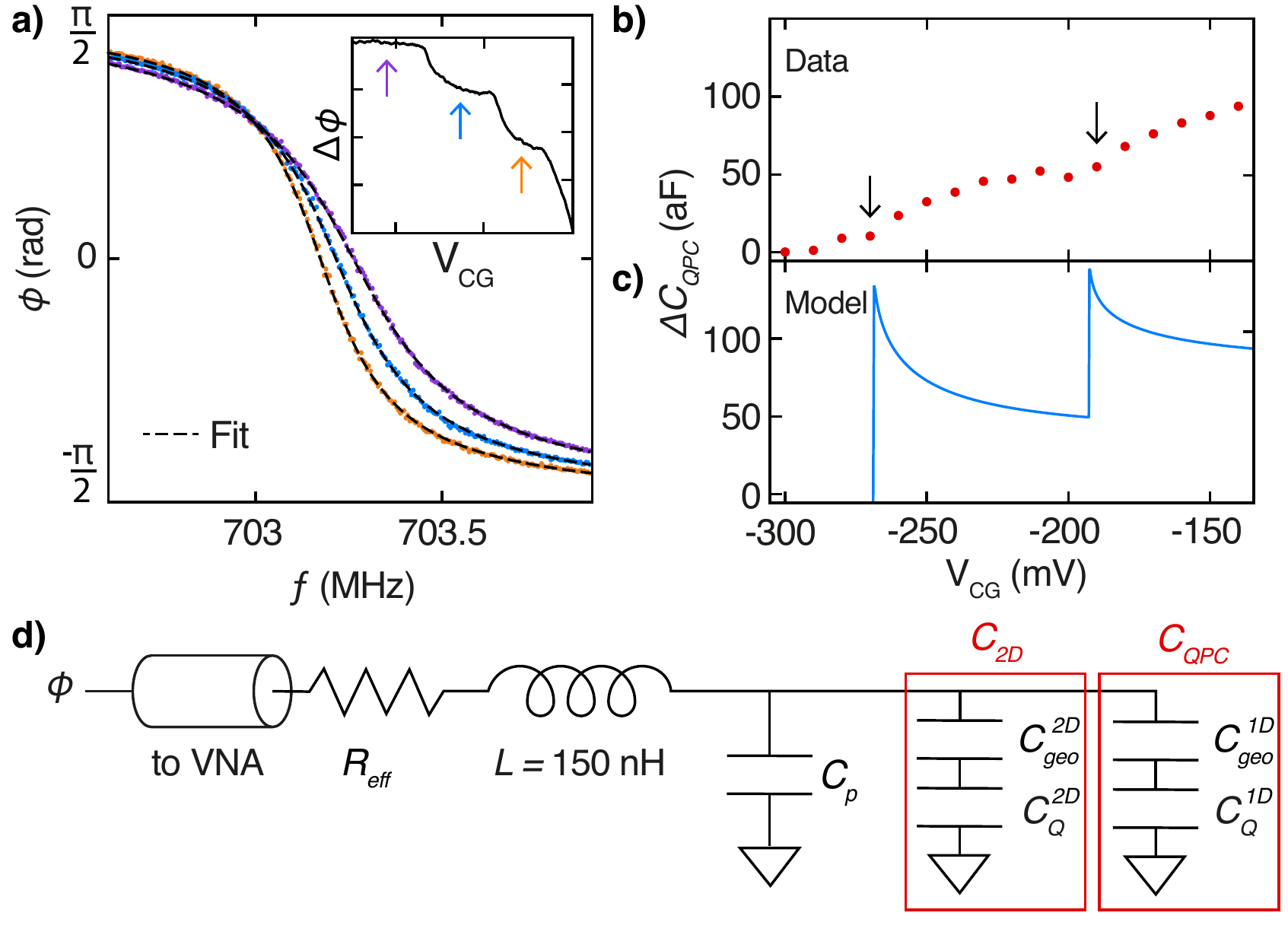}
	\caption{(a) The reflected phase response, $\phi$ for three values of $V_{CG}$. The shift in $f_{0}$ is in due to changing quantum capacitance of the QPC. $\phi$ is fit to a simple LCR circuit model. (b) The change in total capacitance of the system, $\Delta C_{QPC}$, as extracted from the fit to the data in (a), compared with a simple model. We set $\Delta C_{QPC}$ to zero at the value of the total parasitic and geometric capacitance (see main text for explanation). (c) Simplified equivalent circuit of the system showing parasitic capacitance $C_p$, constant capacitance from the depleted 2D region, $C_{2D}$ and a capacitance that changes with gate voltage due to the 1D QPC. Both 2D and 1D regions can be decomposed into geometric $C_{geo}$, and quantum $C_{Q}$ capacitance terms. Throughout the experiment, the input power at device level is $\sim$ -70 dBm.
}
\end{figure}

The change in $\phi$ is due to a dispersive frequency shift of the resonator, which in turn is due to a change in capacitance. We wish to extract this capacitance in order to better understand the origin of the shift and to quantify the sensitivity of the technique. The frequency dependence of $\phi$ near $f_{0}$ is measured for a range of stepped $V_{CG}$ values. Measurements for three such $V_{CG}$ values are shown in Fig. 2(a), with a distinct frequency shift dependent on the number of modes transmitted through the QPC. To extract the total system capacitance, $\phi$ curves for all $V_{CG}$ values are numerically fit to a simple series $LCR$ circuit model, whose total impedance is given by
\begin{equation}
Z_{tot} = j\omega L + \frac{1}{j\omega (C_{p}+C_{2D}+C_{QPC})} + R_{eff}
\end{equation}
where L = 150 nH and $R_{eff}$ = 0.98$Z_{0}$, accounting for a slight impedance mismatch in the circuit. In this equation we separately account for the parasitic capacitance $C_p$, the capacitance that arises from the 2D region under the gate $C_{2D}$, as well as the contribution from the QPC $C_{QPC}$. Setting $V_{SG}$ to a constant value holds $C_{2D}$ also constant, with the resulting change in capacitance with gate voltage due to $\Delta C_{QPC}$, extracted from the fit in Fig. 2(a) and plotted in Fig. 2(b). We note that the dependence of capacitance with gate voltage shows inflections at values of gate voltage that correspond to risers in the conductance staircase [see inset of Fig. 2(a)], suggesting that these inflections (indicated by arrows) come as the edge of each 1D sub-band crosses the chemical potential. 

To gain further insight we estimate $C_{QPC}$ using a simple model, noting that it comprises the series summation of two terms, the geometric capacitance $C_{geo}^{1D}$ between gate and QPC, and the quantum capacitance $C_{Q}^{1D}$, that accounts for the arrangement of charge in the QPC \cite{Gabelli:2006} (see Fig. 2(d) for an equivalent circuit). We first estimate a value of $C_{geo}^{1D}$ $\sim$ 260 aF, using a numerical simulation package \cite{ANSYS}, and then proceed to extract $C_{Q}^{1D}$ using the relation for a 1D system at zero-temperature, given by \cite{Ilani:2006}:
\begin{equation}
C_{Q}^{1D}(\mu) = e^{2}g_{s}l\sum_{j=0}^{n}\sqrt{\frac{m^{*}}{2\hbar^2 \pi^2 (\mu-E_{j})}}
\end{equation}
Where $\mu$ is the chemical potential set by the gate voltage, with lever-arm determined by bias spectroscopy (see Fig. 3). In this equation $E_j$ is the sub-band edge of the $j^{th}$ sub-band, $m^*$ is the effective mass, $l$ is the length of the QPC $\sim$ 200 nm, $g_s$ = 2, is the spin degeneracy ($\hbar$ is Planck's constant). We have ignored broadening of the DOS due to finite temperature, given our approximate electron temperature of 100 mK gives $k_{B}T \sim$ 8 $\mu$eV $\ll$ the 1D sub-band spacing $\sim$ 670  $\mu$eV.  Determining the total change in capacitance of the QPC as, $\Delta 1/C_{QPC} = 1/C_{geo}^{1D} + 1/C_{Q}^{1D}$, Fig. 2(c) plots $\Delta C_{QPC}$ as a function of gate voltage. 

We note that although the model is in broad agreement with our measured data, in that the inflections in the data line-up with the edges of the Van Hove singularities,  there are significant discrepancies. Firstly,  the quasi-1D, rather than truly 1D nature of the QPC is known to lead to broadening of the singularities \cite{Jaksch:2005}. This is particularly relevant in our DGS measurement since changes in the local DOS with gate voltage, away from the 1D constriction, will also likely contribute a background signal that obscures the singularities. Further, we have neglected finite temperature, power broadening from the small oscillating $V_{SG}$ (0.2 mV$_{pp}$), and electron correlation effects \cite{Ilani:2006}.

\begin{figure}
	\includegraphics[width=8.6cm]{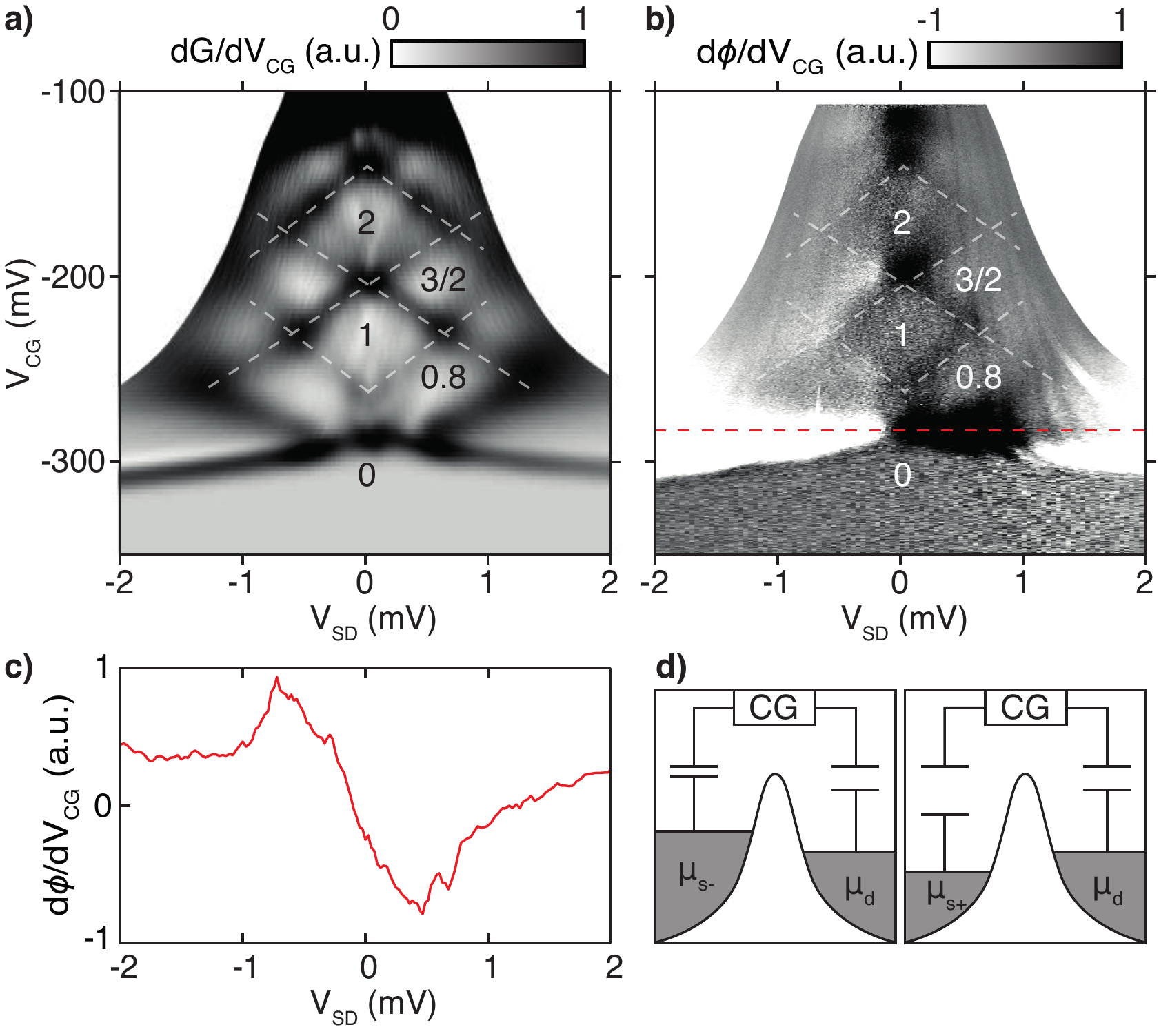}
	\caption{Bias spectroscopy of (a) transconductance and (b) differentiated reflected phase response of the QPC. (c) A slice through $d\phi/dV_{CG}$ at $V_{CG} = -285$ mV (red dashed line), showing a peak at negative bias and trough at positive bias, indicating that the phase response is sensitive to asymmetric application of bias. (d) Cartoon of the source-drain reservoirs relative to one another as bias is changed. The potential of $\mu_{d}$ remains constant, while the potential of $\mu_{s}$ moves above and below it, changing the capacitive coupling between it and the confining gate.}
\end{figure}
We now examine the regime of finite bias, where a source drain voltage can be used to perform spectroscopy of the QPC sub-bands \cite{PhysRevB.44.13549}, noting differences between transport measurements and the DGS technique. It is worth stating that the two techniques are distinct, since with transport the QPC potential is held fixed as the chemical potential either side is varied and probed by a small oscillating voltage. Alternatively in the case of DGS, the gate defined QPC potential is perturbed, oscillating up and down in energy at the resonance frequency as the dc bias is varied. In any case, both the derivative of the ac conductance  $dG/dV_{CG}$ and the derivative of the phase response of the the DGS, $d\phi/dV_{CG}$, exhibit familiar features associated with the 1D sub-bands. Similarity between the measurements can be seen in the structures near zero-bias, and bias values that correspond to the source and drain potentials differing by a sub-band (the so-called finite-bias half-plateaus \cite{PhysRevB.44.13549}). 

A notable difference between transport and DGS is the asymmetry in the features that occur near pinch-off about zero-bias, shown as cut along the red dashed line at $V_{CG} = -285$ mV in Fig. 3(c). Peaks (white) occur at negative bias and troughs (black) occur at positive bias in $d\phi/dV_{CG}$. We attribute this apparent change in the sign of $d\phi/dV_{CG}$ to the asymmetric configuration of source-drain bias relative to the potential of the gate, when a tunnel barrier disconnects the reservoirs either side of the QPC (see Supplementary Material). Whilst the drain potential $\mu_{d}$ is held at ground, the source potential $\mu_{s}$ is raised and lowered, changing the capacitive coupling $C_{2D}$ to the sensing gate, as shown schematically in Fig. 3(d).

\begin{figure}
	\includegraphics[width=8.6cm]{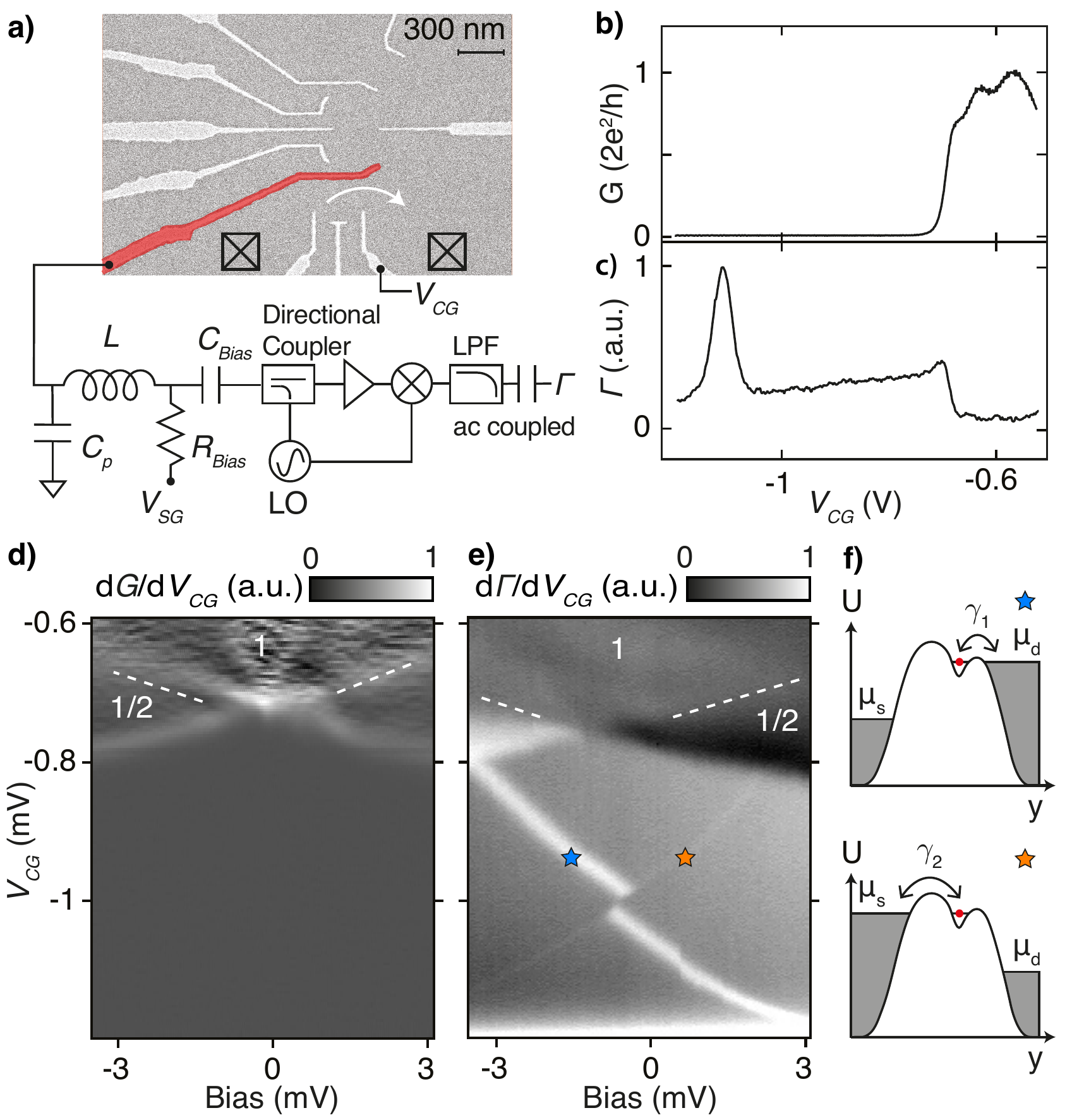}
	\caption{(a) Electron micrograph and experimental setup for a device with an asymmetric QPC geometry, as commonly used for spin qubit experiments. $V_{SG}$ and $V_{CG}$ perform identical functions as before, resonator frequency is at 545.5 MHz. Here we ac-couple the output of the demodulator, realizing a differentiator circuit that avoid the need for numerical differentiation of the data. (b) Conductance and (c) DGS response over the same swept range of $V_{CG}$. A large peak occurs past pinch-off, at approximately $V_{CG}$ =  -1150 mV, visible only in the DGS response. Bias spectroscopy of (d) transconductance and (e) differentiated DGS response, showing the evolution of the peak with bias. There is a stray offset bias of approximately -0.5mV across the device. (f) Cartoon depicting the saddle-point potential, controlled with the gate voltage, relative to the source and drain energy. The data in (e) is consistent with a unintentional quantum dot forming in the QPC past pinch-off, with different tunnel rates $\gamma_{1,2}$ to the reservoirs as indicated by the symbols. }
\end{figure}

Having presented data on a clean QPC that exhibits well-understood behaviour, we now explore a regime where the DGS technique provides new information not easily accessible with transport. Switching to a device with gate pattern commonly used to define a spin qubit based on a double-dot potential, we are able to form a QPC at the tee-intersection of gates shown in  Fig. 4(a). Such a geometry has been used extensively for charge sensing of the double dot when the QPC is biased near pinch-off \cite{Reilly:2007}. The saddle-point potential of the QPC in this gate configuration is likely not as clean as the classic `split-gate' arrangement examined earlier, since the QPC conductance does not exhibit a clear quantized staircase \cite{comment1}. This device has an identical 2DEG depth of 91 nm as our earlier device, but slightly different device parameters of electron density (1.34 $\times 10^{11}$ cm$^{-1}$) and mobility (3.24 $\times 10^{6}$ cm$^{2}$/V s).

Our DGS sensor is configured using the gate shaded red in Fig. 4(a) and  we again compare transport measurements to the response of the DGS, as shown in Fig. 4(b). At the point where the QPC begins to open the DGS response mirrors the step-change seen in conductance, but then strongly contrasts transport in the appearance of a large additional peak, far beyond the gate voltage needed to pinch-off the QPC.

To investigate this anomaly further we perform bias-spectroscopy measurements, plotting conductance and the DGS response side-by-side in Fig. 4(d) and 4(e). Although the features in conductance are not clean, it is possible to make out the first plateau and first sub-band edge before pinch-off. In contrast, DGS clearly shows two diagonal features, one stronger than the other, that cross (or potentially anti-cross) near zero bias. We attribute these features to the presence of an unintentional quantum dot or charge-pocket \cite{Croot:2017}, formed in (or very close to) the QPC saddle potential (see cartoon in Fig. 4(f)). In this sense the diagonal lines correspond to the typical bias-diamonds of Coulomb blockade. Note that in this regime of deep pinch-off, the tunnel barrier is raised well above the potential of the source or drain making it not possible to perform transport measurements. It is unknown if this charge pocket is a genuine trap, or rather a `dark-spot' in the electron wavefunction stemming from the complicated, non-adiabatic constriction imposed by this `T-shaped' gate pattern. Again, we recognise that the DGS technique is sensitive to changes in the local DOS not just in the 1D region of the QPC, but also in the surrounding potential landscape. 

This work was supported by the Centres for Engineered Quantum
Systems (Grants No. CE110001013 and No. CE170100009) and Microsoft Corporation. We thank S. J. Pauka and X. G. Croot for technical assistance and discussions, J. M. Hornibrook for the development of the frequency multiplexing chip, and J. I. Colless for the fabrication of the asymmetric QPC device.
$^{\dagger}$david.reilly@sydney.edu.au

\bibliographystyle{apsrev4-1}

\begin{thebibliography}{35}
\bibitem [{com()}]{comment1}%
  \BibitemOpen
  \href@noop {} {}\bibinfo {note} {This is true not just for the device
  examined here but for many devices we have measured with this gate
  pattern}\BibitemShut {NoStop}%
\end{thebibliography}
%

\subsection{Supplementary Material}
\textbf{Comparing asymmetric and symmetric application of bias.} Using a QPC on a different device, we explore the effect of bias application on the phase response. Rather than using bias spectroscopy, we measure how the response changes as a function of frequency, dependent on the relative potential of the source and drain reservoirs. Asymmetric application, where one reservoir is grounded (as in the text), is shown in Fig. 5 (a), where the phase response has been normalised to that at zero bias. As in the text, peaks and troughs occur about zero bias. In contrast, symmetric application, where the bias is split across the two ohmics, is shown in (b). This time, there is no discernible difference in the response about zero bias. Taking a cut at 408 MHz in (c), we can see a global slope in the phase response as a function of asymmetric bias, due to the constant change in the capacitive coupling to the reservoirs. This does not occur for symmetric bias, whose equivalent schematic is shown in (d), analogous to that presented in Fig 3.(d).
\begin{figure}
	\includegraphics[width=8.6cm]{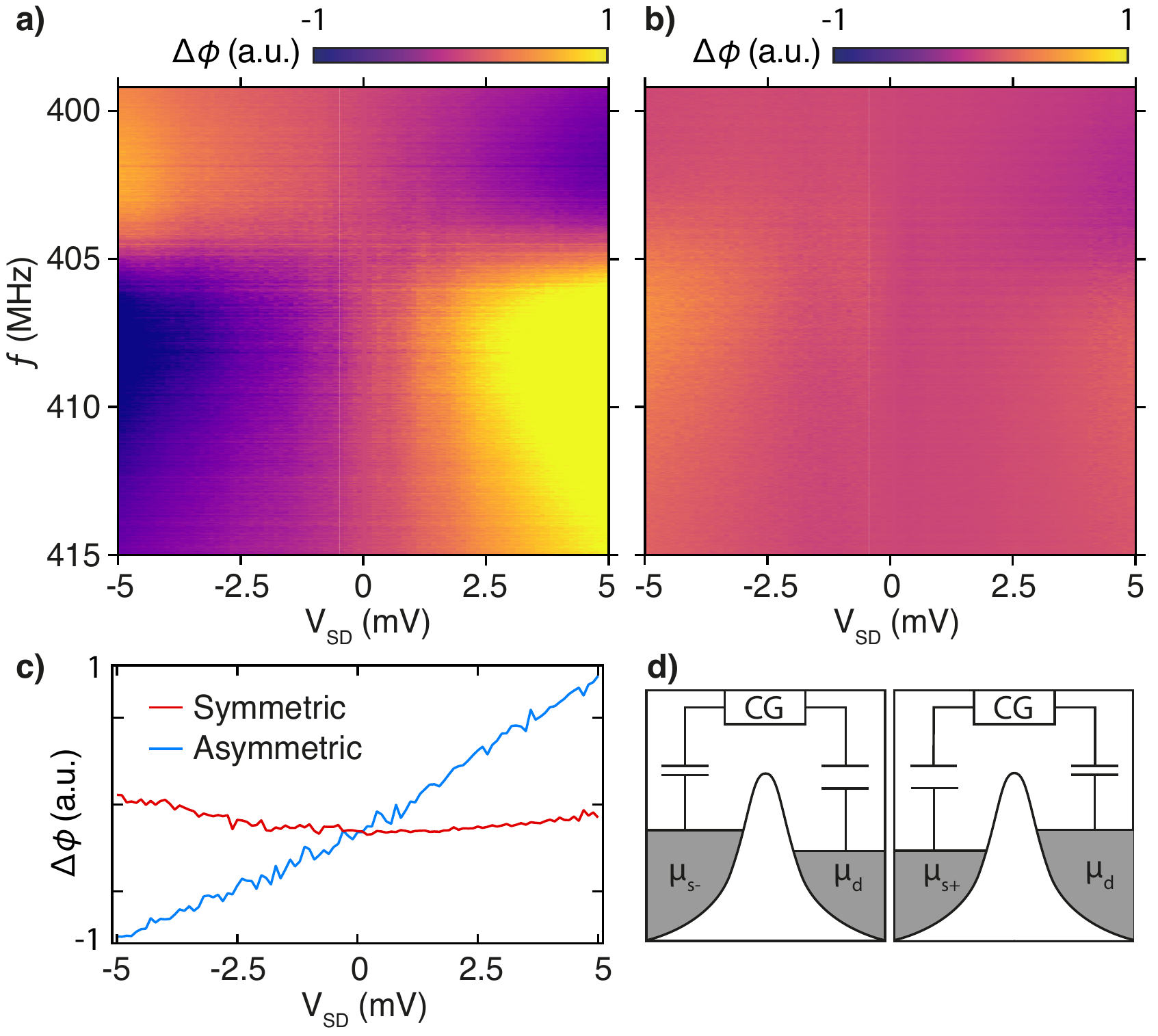}
	\caption{Using a QPC on a seperate device, we explore the phase response a a function of frequency for (a) asymmetric and (b) symmetric applications of bias, normalised to the response at zero bias. The colour scale remains the same between the two plots. Asymmetric application reveals peaks and troughs similar to our measurement, while symmetric application does not. (c) Taking a cut at 408 MHz, we observe a global slope in the phase response about the device for asymmetric application (blue), but not for symmetric application (red). (d) The schematic of reservoir potentials for symmetric bias, analogous to that presented for asymmetric bias in Fig. 3(d).}
\end{figure}
\end{document}